\documentclass[10pt,conference]{IEEEtran}

\usepackage{amsmath,epsfig,amssymb}
\usepackage{bm}
\usepackage{xcolor}
\usepackage{amsthm}
\usepackage[caption=false]{subfig}
\usepackage{cite}
\usepackage[acronym,shortcuts]{glossaries}
\usepackage{lipsum}

\newcommand{\todo}[1]{}
\renewcommand{\todo}[1]{{\color{red} TODO: {#1}}}
\newcommand{\question}[1]{}
\renewcommand{\question}[1]{{\color{red} QUESTION: {#1}}}

\newacronym{GSP}{GSP}{graph signal processing}
\newacronym{GFT}{GFT}{graphical Fourier transform}
\newacronym{DGFT}{DGFT}{digraph GFT}
\newacronym{DV}{DV}{directed variation}
\newacronym{IDV}{IDV}{indefinite directed variation}
\newacronym{CDV}{CDV}{complex directed variation}
\newacronym{TV}{TV}{total variation}
\newacronym{IDGFT}{IDGFT}{indefinite DGFT}
\newacronym{CDGFT}{CDGFT}{complex DGFT}
\newacronym{PCA}{PCA}{principal component analysis}
\newacronym{CS}{CS}{compressive sensing}
\newacronym{FDI}{FDI}{false data injection}

\begin{document}

\pagestyle{empty}
\title{Graph Signal Processing for Infrastructure Resilience: Suitability and
Future Directions}
\author{\IEEEauthorblockN{Kevin Schultz, Marisel Villafa\~{n}e-Delgado,
Elizabeth P.~Reilly, Grace M.~Hwang, Anshu Saksena}
\IEEEauthorblockA{Johns Hopkins University Applied Physics Laboratory\\
11100 Johns Hopkins Road, Laurel, MD, 20723, USA\\
\{kevin.schultz, marisel.villafane-delgado, elizabeth.reilly, grace.hwang, anshu.saksena\}@jhuapl.edu}
}

%\date{\today}

%\pacs{03.65.Wj,02.50.-r}

%
%\thispagestyle{empty}
\maketitle
\begin{abstract}

    Graph signal processing (GSP) is an emerging field developed for analyzing
    signals defined on irregular spatial structures modeled as graphs.  Given
    the considerable literature regarding the resilience of infrastructure
    networks using graph theory, it is not surprising that a number of
    applications of GSP can be found in the resilience domain.
    %
    %GSP techniques can involve several different notions of graphical Fourier
    %transforms (GFTs), but all assume that the choice of GFT imparts a
    %particular spectral structure on the signal.  
    %
    GSP techniques assume that the choice of graphical Fourier transform (GFT)
    imparts a particular spectral structure on the signal of interest.
    We assess a number of power distribution systems with respect to metrics of
    signal structure and identify several correlates to system properties and
    further demonstrate how these metrics relate to performance of some GSP
    techniques.
    We also discuss the feasibility of a data-driven approach that improves
    these metrics and apply it to a water distribution scenario.
    Overall, we find that many of the candidate systems analyzed are properly
    structured in the chosen GFT basis and amenable to GSP techniques, but
    identify considerable variability and nuance that merits future
    investigation.
    %
    %In this work we analyze the underlying critical assumptions of some GSP
    %techniques and assess a range of infrastructure networks with respect to
    %these assumptions.  We then demonstrate the efficacy of these techniques
    %on the networks and provide considerations to their use in practice.
 
\end{abstract}


%\kevin{Everyone check out line 28 or so for custom input commands to track
%questions/comments and additions (vs explicit \todo{}}

\section{Introduction}
Critical infrastructure systems such as power systems and water distribution
are vital components in the safety and security of a nation.
These systems are subject to a wide range of failures and disturbances,
including component failures, natural phenomena such as severe storm events and
earthquakes, and in the increasingly ``wired'' age of information technology,
cyber-events including both malicious attacks and ``benign'' events caused by
unexpected interactions or upgrades
\cite{shinozuka2004resilience,boin2007preparing,duenas2009cascading,
rudner2013cyber,wang2015research}. 
%Furthermore, single failure modes can propagate through the infrastructure
%system (or cross-over into an interdependent systems) in unexpected ways in
%so-called ``cascading'' failures \todo{sprinkle in refs}.
%
Considerable research and investment have been executed to analyze systems in
response to these vulnerabilities,  and to develop novel strategies to manage,
mitigate, and recover any degradation in performance due to them. Collectively,
these efforts have led to the multidisciplinary field of resilience engineering
\cite{hollnagel2006resilience,madni2009towards,patriarca2018resilience} that
seeks to formalize these concepts and apply them to real world situations.

%\grace{do any of the references in 1-9 speak to transportation resilience?
%Because this paper does not specifically address transportation, I would
%suggest deleting transportation from the first sentence}
%
%\kevin{yes, e.g boin2007preparing}

Many of the systems of interest in critical infrastructure have a 
network structure, including lines and buses in a power system, pipes and
junctions in water distribution, or roads and intersections in a transportation
system.  This commonality in network structure has led to extensive
investigation of the resilience of these networks using the language of
mathematical graph theory.  
Graph theory models networks as abstract edges and vertices, and numerous
studies have been performed to show how graph theoretic properties and metrics
relate to resilience in critical infrastructure problems, e.g.,
\cite{holmgren2006using,pagani2013power,pagani2020quantifying,pagano2019water}. 

Motivated by spectral graph theory \cite{chung1997spectral} and algebraic
approaches to signal processing \cite{puschel2008algebraic}, the field of
\ac{GSP} has developed over the past decade to analyze signals defined on
irregular spatial structures modeled as graphs
\cite{shuman2013emerging,sandryhaila2013discrete}.  In the context of
infrastructure systems, example ``graph signals'' include complex phasors on
buses in power systems and hydraulic pressure in water distribution systems.
Several applications of \ac{GSP} to infrastructure systems have appeared in the
literature, including sensor placement problems
\cite{chen2016monitoring,jamei2019phasor,wei2019monitoring}, \ac{FDI}
\cite{ramakrishna2019detection,drayer2019detection}, and general monitoring
problems \cite{jablonski2017graph,hasnat2020detection}.  More broadly, \ac{GSP}
generalizes techniques from signal processing that could find wide utility in
analysis and estimation problems of graph signals in infrastructure
applications, see e.g.,
\cite{zhu2012approximating,chen2015signal,stankovic2019understanding}.

Enthusiasm for this potentially powerful set of tools must be tempered by the
fact that there exists a number of potential approaches to generalizing
Fourier-based techniques to graph signals (i.e., there are numerous ``natural''
ways to define a \ac{GFT}), and each \ac{GSP} technique ultimately relies upon
some assumption or constraint imposed on the graph signals by this \ac{GFT}.
Furthermore, most of the \ac{GSP} power systems literature consider at best a
few different networks, typically of small ($\lesssim$ 100 buses), and no
broad analysis across many systems of different scales has been performed.
To this end, in this paper we analyze particular choices of \ac{GFT} in the
context of a number of different infrastructure systems 
to assess how strongly they meet implied \ac{GSP} assumptions, and then assess
the efficacy of some \ac{GSP} techniques to understand the impact of these
assumptions.  
%\anshu{I don't understand the previous sentence.
%Reword?}\kevin{better now?}
%
This analytical case-study of \ac{GSP} techniques mirror what must be
performed in any practical, real-world application of these techniques to
critical infrastructure.

In the following sections, we first review some graph theoretic and \ac{GSP}
preliminaries and discuss some applications of techniques from the field of
\ac{GSP} to signals defined on infrastructure networks.  We then move to a more
in-depth analysis of the suitability of these techniques for power systems, by
considering power-flow analyses of a large set of diverse power networks and
relate some relevant \ac{GSP} signal metrics to the efficacy of proposed
\ac{GSP} techniques.  Next, we discuss how the physics of water distribution
potentially limits the usage of \ac{GSP} in that domain and offer an attempt to
find a usable \ac{GFT}.
%\grace{Instead of teasing the reader with the potential caveat, is it possible
%to state the nature of the caveat and how water systems differs from power
%systems in the context of using GSP to analyze the resilience of the system? } 
We conclude with additional discussion and future directions for \ac{GSP} in
critical infrastructure.
%directions in the application of \ac{GSP} to infrastructure networks.

%These sorts of  graph theoretic approaches to resilience are primarily
%concerned with understanding how static features of the underlying network
%model relate to resilience concepts, however, they neglect that the network is
%ultimately used for some process, where, for example, edges may have flows that
%meet in the vertices. These processes are constrained in some manner by the
%underlying network, but there may be wide latitude in the dynamics of these
%processes. Consider for example the daily fluctuations in a power or
%transportation grid due to time-varying demand.  The underlying network is
%generally unchanged (except, e.g., concepts like dynamic lane assignment
%in road networks), but presumably certain ``instantaneous'' states of the
%system are inherently more prone to certain forms of threat than
%others. This can be further broken down spatially, as certain regions of the
%system in a given state may be more resilient to a given threat than other
%regions or subsystems.

%We propose that the nascent field of \ac{GSP} can address this apparent gap by
%adapting the fields of Fourier analysis and dynamical systems to processes
%defined on network structures.
%%
%\todo{Marisel: Transition to high level GSP paragraph as evolution}
%\todo{definitely a sentence about extension to  JVT and time varying networks}

\section{Background}

%\kevin{Notation: sets - mathcal, matrices/vectors - bm, scalars - math}

\subsection{Graph Theory Preliminaries}
%
%Mathematically, a graph $\mathcal{G}=(\mathcal{V},\mathcal{E})$ is a
%collection of $N$ vertices $v_k\in\mathcal{V}$ and edges $e_k=(v_\ell,v_m)$.
%Furthermore, associated with the edges may be a set of real or complex-valued
%weights $\{w_k\}$ resulting in a weighted graph. If
%$(v_\ell,v_m)\in\mathcal{E}\implies(v_\ell,v_m)\in\mathcal{E}$ (and any
%associated weights are equal), the graph is said to be \textit{undirected},
%otherwise the graph is said to be \textit{directed}.  A convenient way to
%represent a graph is through the adjacency matrix $A$\todo{\dots}

Mathematically, a \textit{(undirected) graph}
$\mathcal{G}=(\mathcal{V},\mathcal{E})$ is defined as a collection of a set of
vertices (sometimes called nodes) $\mathcal{V}=\{v_1,\dots,v_N\}$ and a set of
edges $\mathcal{E}=\{\{u,v\}:u,v\in\mathcal{V}\}$.  $\mathcal{G}$ is said to
have $N$ vertices and $|E(G)|=M$ edges.  If $\{u,v\} \in \mathcal{E}$, then $u$
and $v$ are said to be \textit{adjacent}.  Given a vertex $v \in \mathcal{V}$,
the \textit{degree} of $v$ is $d(v)=|\{u:\{u,v\} \in \mathcal{E}\}|$, or the
number of nodes adjacent to $v$.  A \textit{directed graph} is a graph where
the edges are ordered pairs of vertices rather than sets. In other words,
$\mathcal{E}=\{(u,v):u,v \in \mathcal{V}\}$. The \textit{head} of edge $(u,v)$
is $v$ and the \textit{tail} is $u$.  Graphs can also have vertex and/or edge
attributes associated with them.  In the context of this paper, we are
interested in edge weighted graphs, where each edge $e_k$ in $\mathcal{E}$ has
an associated weight $w_k$, or vertex weighted graphs, where each vertex $v_k$
has an associated weight $w_k$.  
%A graph is said to be \textit{connected} if, for every pair of vertices, there
%exists an undirected path between the two vertices. A directed graph is said
%to be \textit{strongly connected} if, for every pair of vertices $u$, $v$,
%there exists a directed path from $u$ to $v$ and vice versa.  

The \textit{adjacency matrix}, $\bm{A}$, is one convenient way of representing
the structure of a graph where element $A_{k\ell}=1$ (or the edge weight for edge
weighted graphs) if $v_k$ is adjacent to $v_\ell$ and $A_{k\ell}=0$ otherwise.
Note that $\bm{A}$ is symmetric if $\mathcal{G}$ is undirected, but $\bm{A}$
need not be symmetric if $\mathcal{G}$ is directed. 
%but not necessarily for directed graphs.
The \textit{incidence matrix}, $\bm{B}$, is another way of storing structural
information.  For an undirected graph, it is an $N\times M$ matrix where
element $B_{k\ell}=1$ if vertex $v_k$ is an element of edge $e_{\ell}$ and
$B_{k\ell}=0$ otherwise. For a directed graph, element $B_{k\ell}=1$ if $v_k$
is the head of $e_{\ell}$, and $B_{k\ell}$ is -1 if it is the tail.  For edge
weighted graphs, the $\pm1$ entries of $B$ are replaced with $\pm w_\ell$.  For
undirected graphs, the \textit{Laplacian matrix}, $\bm{L}$, can be calculated
from the adjacency matrix.  Specifically, it is defined as
$\bm{L}=\bm{D}-\bm{A}$ where $\bm{D}$ is a diagonal matrix such that
$D_{kk}=d(v_k)$ for vertex $k$ and $\bm{A}$ is the adjacency matrix.  It is
common to consider the \textit{spectrum}, or the set of \textit{eigenvalues}
$\lambda_k$ of $\bm{L}$, to understand certain properties of the graph such as
clustering, and for positive weighted, undirected graphs, $\lambda_k\geq0$
\cite{chung1997spectral}. For directed graphs or graphs with negative or
complex-valued weights, we will use the underlying positive, undirected
adjacency matrix $\bm{A}^{(|u|)}$ where
$A^{(|u|)}_{k\ell}=\max\{|A_{k\ell}|,|A_{\ell k}|\}$
\cite{shafipour2018directed,schultz2020graph} to derive a corresponding
underlying $\bm{L}^{(|u|)}$.  Other generalizations of graph Laplacians exist
in the literature, as well \cite{chung2005laplacians}.

\subsection{Graph Signal Processing}

\begin{figure}[h]

    \centering

    \includegraphics[width=.9\columnwidth]{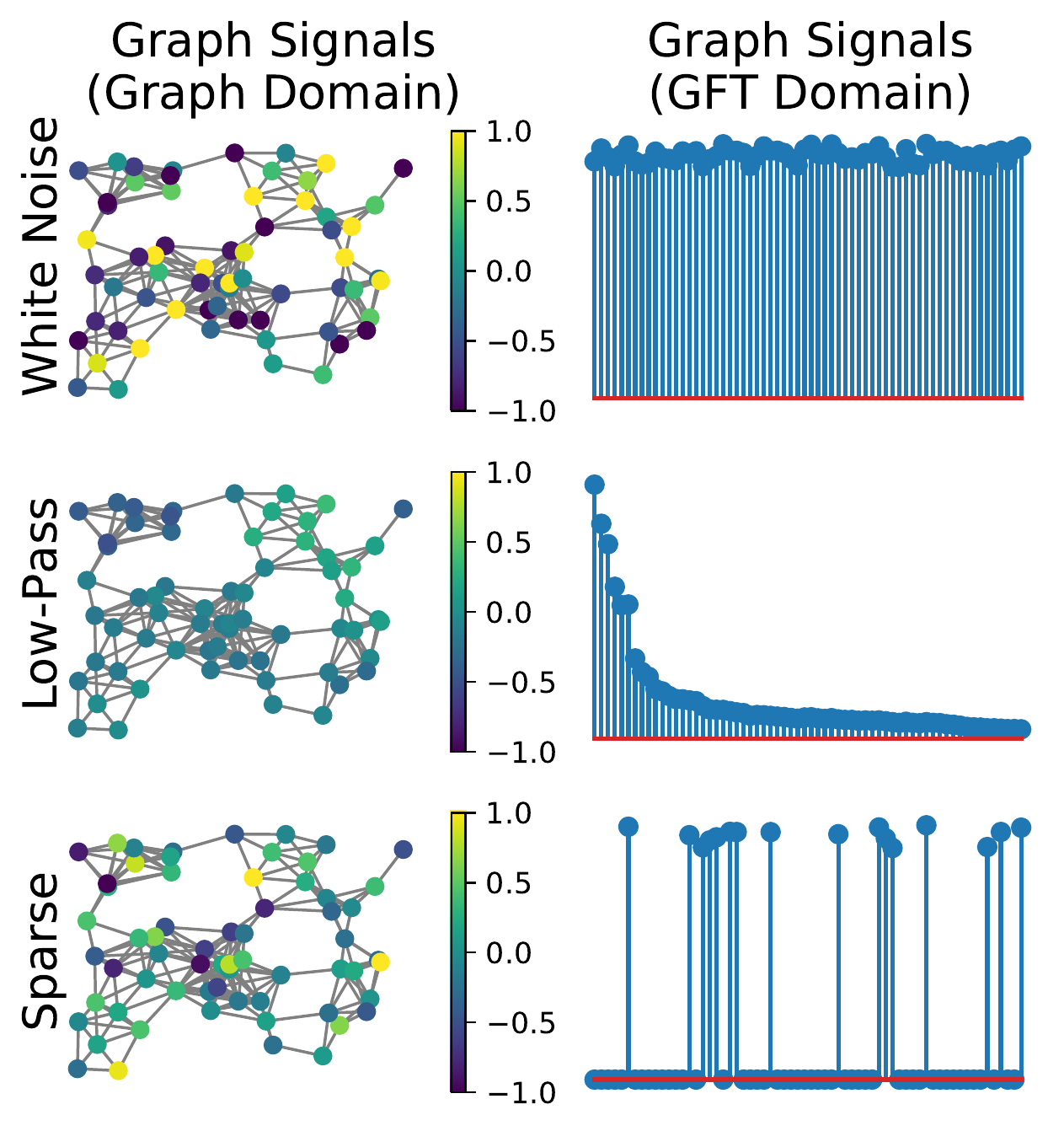}

    %\caption{Efficacy of denoising approaches as a function of noiseless
    %signal low-pass compressibility and \ac{TV}. Note the $x$-axis is used for
    %both quantities.}
    \caption{Example graph signals (left column) and their representations in
    the \ac{GFT} domain (right column). White noise signals are random on the graph
    and are uniform in the \ac{GFT} domain. Low-pass graph signals vary
    smoothly with respect to the graph topology, and are concentrated in the
    low-frequency graph harmonics.  Sparse signals can have high-frequency
    graph components and appear seemingly random in the graph domain, but are
    highly structured in the \ac{GFT} domain.}
    
    \label{fig:example}

\end{figure}

\ac{GSP} is a field that has emerged over the past decade and aims to
generalize signal processing techniques to signals defined over graphs
\cite{sandryhaila2013discrete,shuman2013emerging}. 
In classical signal processing, signals live in a Euclidean space, as for
example signals defined over time, images, and video.  
In \ac{GSP}, the nodes of a graph define the domain of the so-called graph
signal 
%\todo{add an illustrative figure?}.
Formally, given a graph $\mathcal{G}=(\mathcal{V}, \mathcal{E})$, a graph
signal $\bm{f}:\mathcal{V}\to\mathbb{V}$ is a function defined on the vertices
of $\mathcal{G}$ that takes values in some vector space $\mathbb{V}$, typically
$\mathbb{R}^N$, although $\mathbb{C}^N$ is a natural domain for power
applications.  For notational convenience we will denote graph functions
$\bm{f}(v_k)$ by $f_{k}$ emphasizing that the function can indeed be viewed as
a vector ``attached'' to a graph.
Several analysis techniques and transforms from classical signal processing
have been adapted to signals defined over a graph.  Key to these ideas is the
notion of an $N\times N$ (possibly complex) \ac{GFT} matrix $\bm{U}$ that
defines the \ac{GFT} $\tilde{\bm{f}}=\bm{U}^\dagger\bm{f}$, where $\dagger$
denotes conjugate-transposition. When $\bm{U}$ is unitary, its columns
$\bm{U}_k$ are orthonormal and there is a natural inverse \ac{GFT} defined by
$\bm{f}=\bm{U}\tilde{\bm{f}}$.  This orthogonality captures much of the
original intuition behind the discrete time Fourier transform and is exploited
throughout \ac{GSP} applications.  Figure~\ref{fig:example} shows some example
graph signals and their representations in the \ac{GFT} domain.

%\todo{This definition will
%fit better if moved to the graph background. Keeping it here in the meantime}

There are many ways to define a \ac{GFT} from a given graph $\mathcal{G}$, some
of which do not necessarily produce a unitary transform.  Some approaches more
motivated by algebraic signal processing use $\bm{A}$ directly, while others
exploit spectral graph theoretic motivation of Laplacians
\cite{sandryhaila2013discrete,shuman2013emerging,
shafipour2018directed,schultz2020graph}.  Here, we will restrict ourselves to
Laplacian-derived unitary \ac{GFT}s.
For undirected graphs with positive edges, the Laplacian $\bm{L}$ admits the
eigendecomposition $\bm{L} = \bm{U}\bm{\Lambda}\bm{U}^T$, where $\bm{U}$ is the
matrix with its eigenvectors $\bm{U}_k$ along its columns and $\bm{\Lambda}$ is
a diagonal matrix of the eigenvalues, $0 = \lambda_1 \leq \cdots \leq
\lambda_N$.  
%\grace{is permits a better verb than admits or is admits a verb that
%mathematicians prefer?}\kevin{admits is used in math}
%
For general graphs, $\bm{L}^{(|u|)}$ can be used to define similar transforms
\cite{shafipour2018directed,schultz2020graph}.
%For a graph signal $\bm{f}\in \mathbb{R}^N$ \ac{GFT} 
%
%\todo{add more about the motivation from differential calculus and expand on
%why the eigenvectors of the Laplacian as basis} is defined as $\tilde{\bm{f}}
%= \bm{U}^{T}\bm{f}$.
%
The eigenvalues $\lambda_k$ of $\bm{L}$ are used to define a notion of
frequency for the graph harmonics defined by eigenvectors $\bm{U}_k$.  Unlike
standard Fourier analysis, these ``frequencies'' are not uniformly spaced, but
they can be ordered.  In particular, the harmonics corresponding to
$\lambda_k=0$ correspond to the average value for that connected component.
In the \ac{GSP} context, the \ac{GFT} provides an illustration of alignment or
smoothness between the graph signals and their adjacent graph edges in much the
same way that frequency content of a signal in standard Fourier analysis does.
%\todo{clarify/expand}
An important measure is the \ac{TV} of a graph signal, defined as $TV(\bm{f}) =
\bm{f}^\dagger\bm{L}\bm{f} = \sum_{k,\ell=1, \ell>k}^{N} A_{k\ell}|f_k -
f_\ell|^2 $.  In the context of the eigendecomposition of $\bm{L}$, \ac{TV} is
the average frequency weighted by the \ac{GFT} power spectrum of $\bm{f}$.

%\todo{expand}

Another important concept in \ac{GSP} is that of a graph filter, where a graph
signal $\bm{f}$ is modified in the \ac{GFT} domain and transformed back into
the graph signal domain, which can be expressed as
$\hat{\bm{f}}=\bm{U}\text{diag}(\bm{h})\bm{U}^\dagger\bm{f}$ where the vector
$\bm{h}$ re-scales (and/or phase-shifts) the signal energy in each \ac{GFT} harmonic, in analogy with
window methods in classical signal processing.  For example
$\bm{h}=[1,1,\cdots,1,0,0,\cdots 0]^\top$ defines a low-pass graph filter,
since all the harmonics above some graph frequency cutoff are set to zero.

\section{Applications in Resilience}\label{sec:appl}

Despite its relative nascence, the field of \ac{GSP} has rapidly generalized a
number of techniques from ``standard'' signal processing to functions defined
on the nodes of a graph, many of which appear to be suitable for infrastructure
applications.  These include signal denoising \cite{chen2015signal} and more
generally filtering of noisy signal $\breve{\bm{f}}=\bm{f}+\bm{n}$, where
$\bm{n}$ is some noise graph process.  Much like classical signal processing,
the idea is to apply some graph filter $\bm{h}$ that removes as much of
$\bm{n}$ while preserving $\bm{f}$, or otherwise extracting spectral components
of $\breve{\bm{f}}$.  Similarly, when there is both noise and an external
\ac{FDI} signal $\bm{e}$ (so $\breve{\bm{f}}=\bm{f}+\bm{n}+\bm{e}$), a graph
filter $\bm{h}$ that extracts the expected spectral content of $\bm{f}$ and
$\bm{n}$ can be used to identify potential attacks $\bm{e}$
\cite{ramakrishna2019detection,drayer2019detection}.  When $\bm{f}$ is expected
to be sparse (i.e., $\tilde{\bm{f}}$ is concentrated in a few harmonics), this
signal structure can be exploited to estimate $\bm{f}$ using only the values
at a few vertices \cite{zhu2012approximating,chen2015signal}.

%sparse reconstruction
%\cite{zhu2012approximating,chen2015signal}, sensor placement
%\cite{chen2016monitoring,jamei2019phasor}, and false data injection attacks and
%detection \cite{ramakrishna2019detection,drayer2019detection}.
%
%We will now briefly introduce these problems and the general intuition behind
%their \ac{GSP} motivation as many of these problems are quite broad and have
%considerable literature discussing various nuances and approaches involving
%different network models, graph shift operators and Fourier transforms, and
%other variations on the general problem. 

%\todo{briefly/notionally define the above problems}

\subsection{GSP as a signal model}\label{sec:sig_model}
The efficacy of any of the aforementioned \ac{GSP} techniques is dependent upon
the choice of the \ac{GFT} used. It is assumed that the class of
expected or typical signals will have some general structure with respect to
the chosen \ac{GFT}, typically ``low-pass'' or spectrally sparse in the chosen
\ac{GFT} domain (see Figure~\ref{fig:example}).  If, for example, the typical
graph signals are uniformly spread (i.e., ``white'' noise) with respect to a
given \ac{GFT}, then these GSP techniques will not be as
effective.  For example, \ac{FDI} detection would basically devolve into
detection based on individual nodes, rather than the entire signal.
%Such an attack can be assessed based on two tests: 1) total injected signal
%energy (measured across the entire \ac{GFT} domain and compared to the typical
%signal energy) or 2) comparisons (e.g., log-likelihood tests) at typical
%values at each node. 
Such a test would be less sensitive than under the low-pass model, which
compares signal energy at the high frequency harmonics in the \ac{GFT} domain
with expected values based on typical signal smoothness.  This improves the
detection rate of the test by exploiting the overall assumption of signal
smoothness. In this sense, the assumption of a particular signal model in a
specific \ac{GFT} domain needs to be evaluated for the specific network and
``typical'' signals defined on it.

Many of these problems have been addressed under alternative approaches
and signal models, for example using \ac{CS} and sparse optimization approaches
\cite{liu2014detecting} or data-driven approaches based on e.g., \ac{PCA}
\cite{valenzuela2012real}.  
Sparsity inducing bases for \ac{CS} models and the principal component vectors
can both be viewed as unitary matrices that define transforms that induce a
signal structure, much like a \ac{GFT}.
Comparison to these two classes of approaches elucidates what the \ac{GSP}
framework is actually assuming, namely, that the network structure and some
underlying physical model induces a particular signal structure that can be
inferred using only the network structure.  This is a stronger assumption than
the \ac{PCA} approach, which exploits the existence of signal structure
determined empirically, and on par with many \ac{CS} approaches where the
sparse basis is known \textit{a priori}.
%
%
%
%In \ac{CS} \todo{relate GSP to sparsity inducing basis in CS}
%
%Similarly, \ac{PCA} makes an assumption that the data lies in
%reduced-dimensional subspace (possibly perturbed by noise), in this case
%determined by dominant singular vectors determined empirically by some
%previously obtained data. Typically, this data is assumed to be Gaussian
%distributed, and a number of statistical tests can be devised to exploit this
%to perform different estimation or learning tasks.  In particular, if we let
%$U=[S, S_\perp]$ be the unitary matrix defined by concatenating span of the
%assumed signal subspace $S$, and its null-space $S_\perp$, then $U$ can be
%used in much the same way as a \ac{GFT} operator, and the assumptions of
%\ac{PCA} imply that the data should be low-pass with respect to this operator.
%
This is not meant to be taken as a criticism of \ac{GSP} approaches, as it
might initially seem that we are advocating for more direct data-driven
approaches that more rigidly (at least empirically) induce the signal structure
assumptions exploited in the various techniques above. Instead, these \ac{GSP}
motivated approaches attempt to capture something ``universal'' and physically
motivated about a particular class of infrastructure systems.
%\todo{something about unexploited physical knowledge and constraints, broader
%prior, uncertainty type stuff}.

\section{Power Systems}\label{sec:power}
A number of GSP applications can be found in the domain of power systems, at
least partially due to the rich set of potential graph models to use for
exploitable signal structure. Modeling the graph as unweighted uses only
connectivity and ignores all information about the lines
themselves, and due to this, appears to be unused in this domain.  In
\cite{hasnat2020detection}  the inverse of the length of each line in the power
system is used to weight the edges.  The decoupled (DC) power flow assumptions
imply a graph weighted by the inverse of the reactance (and a graph signal of
phases). The above discussion highlights a combinatorial issue that needs to be
resolved in order to effectively apply \ac{GSP} techniques, as one must find
both a (weighted) graphical structure and a corresponding \ac{GFT} operator
that effectively induces the desired exploitable signal structure.  Just
because one combination of network weights and \ac{GFT} does not appear to
induce the required structure does not rule out the existence of a useful
model.  

Here, we consider weights motivated by (idealized) alternating current (AC)
power flow by defining the admittance
$y_{k,\ell}=(r_{k,\ell}+jx_{k,\ell})^{-1}$ of a line in terms of the resistance
$r_{k,\ell}$ and reactance $x_{k,\ell}$.
The current flow $i_{k,l}$ through a line is related to the admittance %
%\begin{equation} % y_{k,\ell} = \frac{1}{r_{k,\ell}+jx_{k,\ell}} %
%\end{equation}
%
and the bus voltages $V_k$ and $V_\ell$ via
$i_{k,\ell}=y_{k,\ell}(V_k-V_\ell)$. 
%Solving these equations in the presence of generators and loads is the AC
%power flow problem. 
By assigning a fixed (yet arbitrary) direction to each line, this set of
equations can be used to define a directed incidence matrix $\bm{B}_{AC}$ that
relates the vector of currents on each edge $\bm{i}$ to the vector of voltages
$\bm{V}$ by $\bm{i}=\bm{B}_{AC}^\dagger\bm{V}$.

In what follows, we restrict ourselves to analyzing the complex vector of bus
voltages, $\bm{V}$, using the AC power flow network structure under the
\ac{GFT} defined by $\bm{L}^{(|u|)}$ for $\bm{B}_{AC}$.  Using this as the
basis for our \ac{GSP} analysis, we consider AC power-flow simulations computed
by \texttt{MATPOWER} \cite{zimmerman2010matpower} using the example networks
provided by that package. We restrict our analyses to networks with bus counts
less than 4096 to produce a set of default bus voltages to serve as graph
signals.  This set of 43 networks spans 4--3374 buses,
and includes both IEEE test cases (14, 24, 30, 57, 118, 145, 300 bus cases)
and systems  modeling complex, real world networks from Texas, France, and
Poland, see \cite{zimmerman2010matpower} source code.% for more information.

\subsection{Suitability of \ac{GSP} Approaches}
As discussed in Sec.~\ref{sec:sig_model}, a common assumption about the signal
structure for \ac{GSP} applications is that the signal is in some sense
compressible, which means that the signal is well approximated by a few
\ac{GFT} harmonics. Let us first define three metrics to illustrate when
\ac{GSP} approaches may be suitable for our analysis problem of choice. The
first metric we consider is low-pass compressibility, which is defined by the
number of consecutive low-frequency  graph Fourier harmonics required to
capture some fraction of the overall signal power, here 90\% and 99.9\%.
Low-pass compressible graph signals would be ideal candidates for denoising via
a low-pass graph filter.  Additionally, low-pass compressibility is exploited
in \ac{FDI} detection approaches; for example the 99.9\% threshold was used in
\cite{hasnat2020detection} as a cutoff for high-pass graph filter to detect
\ac{FDI}.
The second metric we consider is  general compressibility, which is defined by
the minimum number of harmonics required to capture the desired fraction of
energy, without the low-pass constraint. Thus, general compressibility is a
metric of signal sparsity in the \ac{GFT} domain. Graph signals that are sparse
can be reconstructed from fewer measurements, and this can be exploited to
reduce the number of sensors required to monitor the system.  The difference
between low-pass compressibility and compressibility (i.e., sparsity) can be
seen in Figure~\ref{fig:example}.
The third metric we use is total variation $TV(\bm{V})$, an alternative notion
of signal smoothness that is commonly encountered in the \ac{GSP} literature as
a regularizer in optimization problems. To normalize the above metrics, we
divide the notions of compressibility by the number of buses to compute
compressibility ratios, and we normalize $TV(\bm{V})$ by dividing by
$||\bm{V}||^2$ for each graph signal.
%Third, while low-pass compressibility captures a notion of signal smoothness,
%another common notion of signal smoothness encountered in the literature is
%\ac{TV}, defined here by
%%
%\begin{equation} % TV(V)=\frac{V^\dagger \mathcal{L}_{|A|}V}{||V||^2}\,.  %
%\end{equation}

%\todo{Kevin: Here, I would suggest adding a sentence to tell the reader what
%type of signal would be compatible with GSP analysis before diving into the
%results. } Ideally, highly compressible signals (compressibility ratio $<$
%0.1) would be conducive to \ac{GSP} analysis. Similarly, total variation that
%scales monotonically with signal-to-noise ratio would also be appropriate for
%\ac{GSP} analysis. \todo{Kevin/Marisel: are these preceding sentences true?}

\begin{figure}[h]

    \centering

    \includegraphics[width=.9\columnwidth]{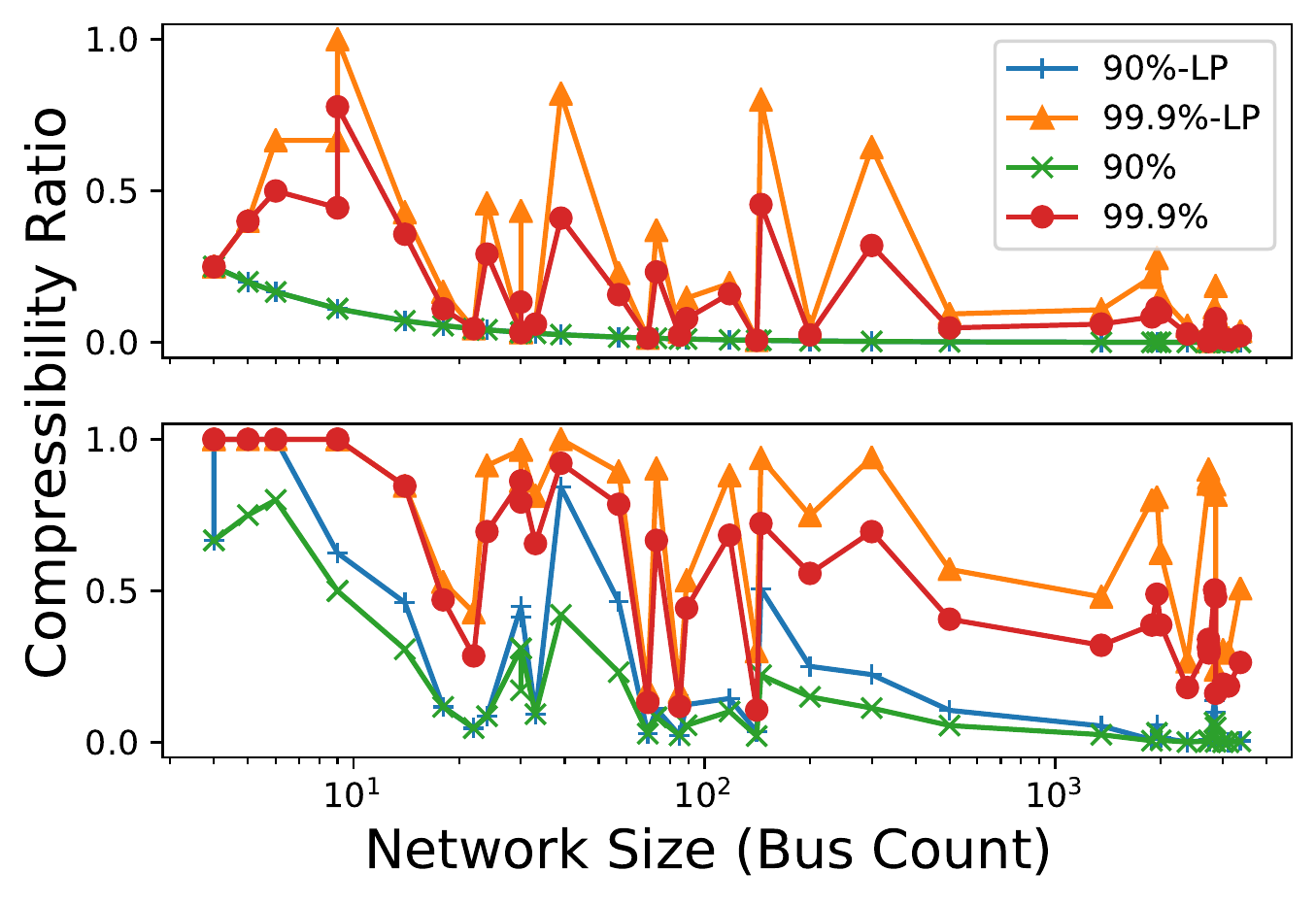}

    \caption{Low-pass (LP) and general compressibility ratios of bus voltages
    at 90\% and 99.9\% thresholds.  \textbf{Top}: Compressibility ratios of the
    complete signal. \textbf{Bottom}: Compressibility ratios with the signal
    average removed.  } \label{fig:compressibility_ratios}

\end{figure}

For each network in our \texttt{MATPOWER} test set, we used the default AC
power-flow result to assess its low-pass and general compressibility.
Figure~\ref{fig:compressibility_ratios}~(top) shows the compressibility ratios
for each graph signal. This indicates that many of these signals are reasonably
compressible (especially at the 90\% threshold) and are generally inversely
correlated with network size ($r_s=-0.99$, $p=3.9\times10^{-38}$ for 90\% and
$r_s=-0.53$, $p=2.3\times10^{-4}$ for 99.9\% using Spearman's rank correlation
coefficient $r_s$ \cite{myers2010research}). The general compressibility
notions are similarly correlated.
%\todo{general compressibility} \todo{Kevin: Grace finds the next sentence
%confusing because we never show the data on the 0th harmonic; is it common
%knowledge to your intended audience? I added (not shown) to the sentence in an
%attempt to help the reader along.}% 
It turns out, however, that these bus voltages are dominated by the average
voltage across the entire bus (c.f., the DC approximation) which is captured by
the $\lambda=0$ graph harmonic. When we look at the compressibility of only the
remaining harmonics, i.e., the perturbations from the signal mean, we see that
the signals are far less compressible, see
Figure~\ref{fig:compressibility_ratios}~(bottom).  The discrepancy between
these two panels indicates that sparse reconstruction (equivalently, sparse
sensor placement) problems that care only about perturbations from the mean
will require more measurements to reconstruct to the same level of error, as
the compressibility ratio of a signal essentially determines the number of
measurements required to reconstruct the signal at a given error threshold.
%
%
%The discrepancy between these two panels indicates that the efficacy (in terms
%of reconstruction error) of sparse reconstruction (equivalently, sparse sensor
%placement problems) is dependent upon the relative importance (for a given
%application) of the perturbation of the voltage signal from the average across
%the network, because the compressibility ratio of a signal essentially
%determines the number of measurements required to reconstruct the signal at a
%given error threshold.  \anshu{Reword this last sentence?  Is sparse
%reconstructability really dependent on relative importance, or is it just that
%the compressibility metric is, suggesting that whether to include DC should be
%carefully considered?  Is the data being shown here actually compressible? And
%can the plots be line graphs rather than scatter?  Harder to see as scatter}
%\todo{Kevin: the connection to optimal sensor placement as an equivalence to
%sparse reconstruction seems like a surprise. Is Figure 1 an analysis on sensor
%placement? Why does the discrepancy revealed by low-pass and general
%compressibility suggest that we have a solution to optimal sensor placement?
%As a reader, my take-home message would be that I should always remove the
%zeroth harmonic, and that general compressibility metric produces a lower
%compressibility ratio which is a good thing. I wonder if it should be
%explicitly stated that networks size with compressibility ratio=1 is not
%suitable for GSP analysis ?  }
%

\begin{figure}[h]

    \centering

    \includegraphics[width=.9\columnwidth]{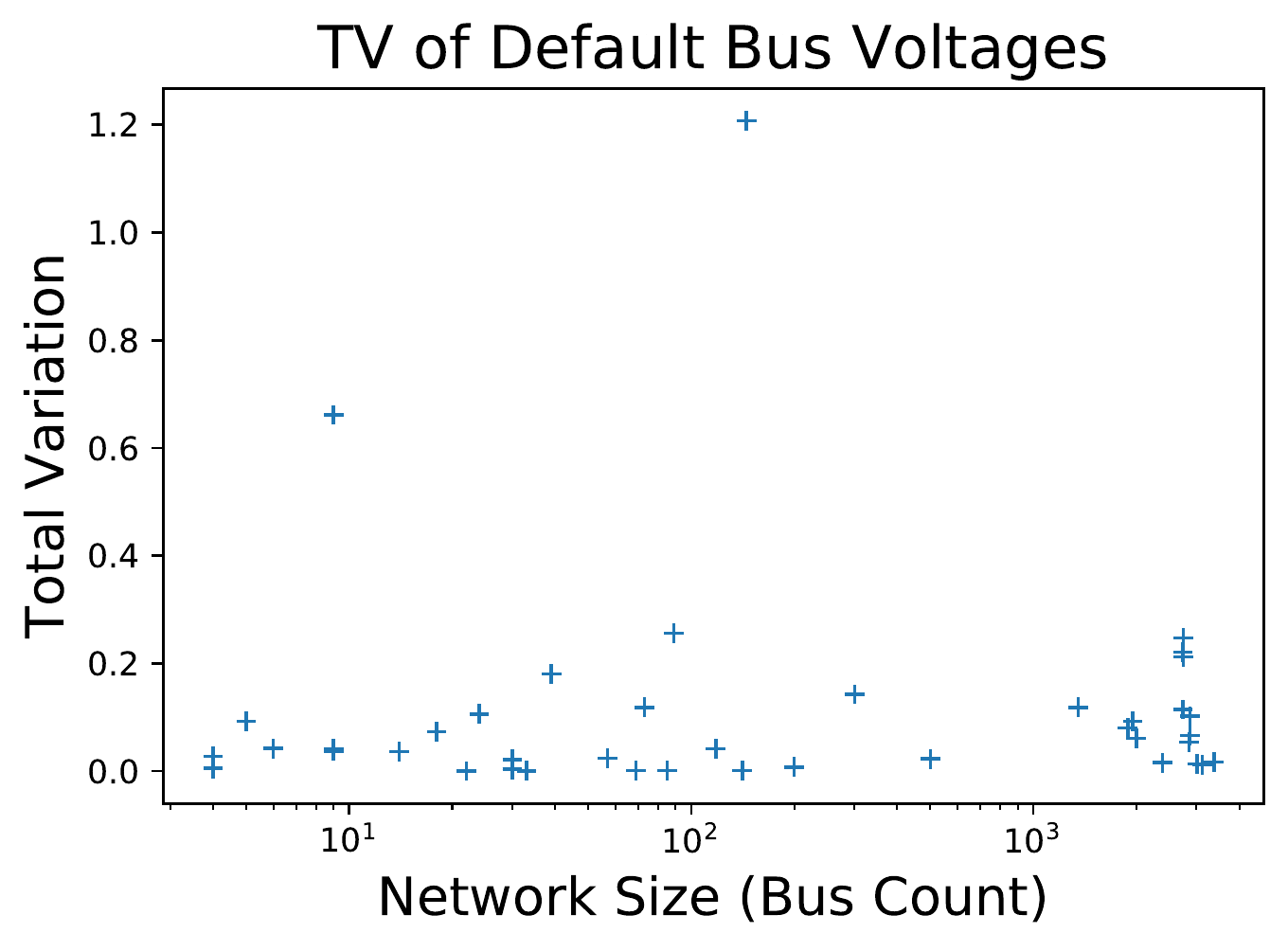}

    \caption{Total Variation of AC power flow-computed bus voltages compared to
    network size.}
    \label{fig:TV_by_size}

\end{figure}

Figure~\ref{fig:TV_by_size} shows the normalized \ac{TV} for each of the
MATPOWER test cases. 
Unlike the low-pass and general compressibility which was inversely correlated
with network size, here we see that \ac{TV} is not particularly correlated with
network size ($r_s=0.19$, $p=0.22$).  The largest outlier in
Figure~\ref{fig:TV_by_size} corresponds to the IEEE 145 bus, 50 generator
dynamic test case, in which over a third of the buses have active generation
capability. Motivated by this observation, we performed a correlation analysis
that related percentage of buses with active generation to normalized \ac{TV}
and we found a moderately strong correlation ($r_s=0.44$,
$p=2.9\times10^{-3}$). Since distributed power generation capability injects
extra ``degrees of freedom'' in the power flow computations, it is not
surprising that this produces signals that are in some sense less smooth.
Performing a similar analysis indicate that compressibility at the 99.9\%
threshold is not statistically significantly correlated ($r_s=-0.17$, $p=0.25$,
for low-pass, $r_s=-0.28$, $p=0.07$ for general), but that 90\% compressibility
is (both $r_s=-0.83$, $p=3.38\times10^{-12}$), albeit less strongly correlated
than they are to network size.

%\todo{Kevin/Marisel: Now I'm wondering what to do with TV? It doesn't look meaningful as a function of network size. When would a practitioner of GSP choose to use the TV metric over the LP- compressibility metric? We just showed how LP-compressibility was obscured by the zeroth harmonic. Leaves me with the impression that both metrics are less than ideal. Grace}

%\todo{Anyone want to make an illustration showing low pass vs general
%compressibility counts, maybe a TV/smoothness one would be good, too -- maybe
%goes up in signal model? Grace may try} 

\subsection{Impact on \ac{GSP} Techniques} 
The previous section focused on how well the set of candidate systems and
signals met some common \ac{GSP}-motivated signal models.  In this section, we
analyze some \ac{GSP} techniques applied to the same set of systems and signals
to assess the impacts of the metrics from the previous section on these
approaches.
Applications related to sparse reconstruction and approximation, such as the
optimal sensor placement problem have already been discussed at a notional
level in the context of Figure~\ref{fig:compressibility_ratios}, so we will
instead focus on estimating the true bus voltages from those corrupted by noise
(the denoising problem) and \ac{FDI}s.

\begin{figure}[h]

    \centering

    \includegraphics[width=.9\columnwidth]{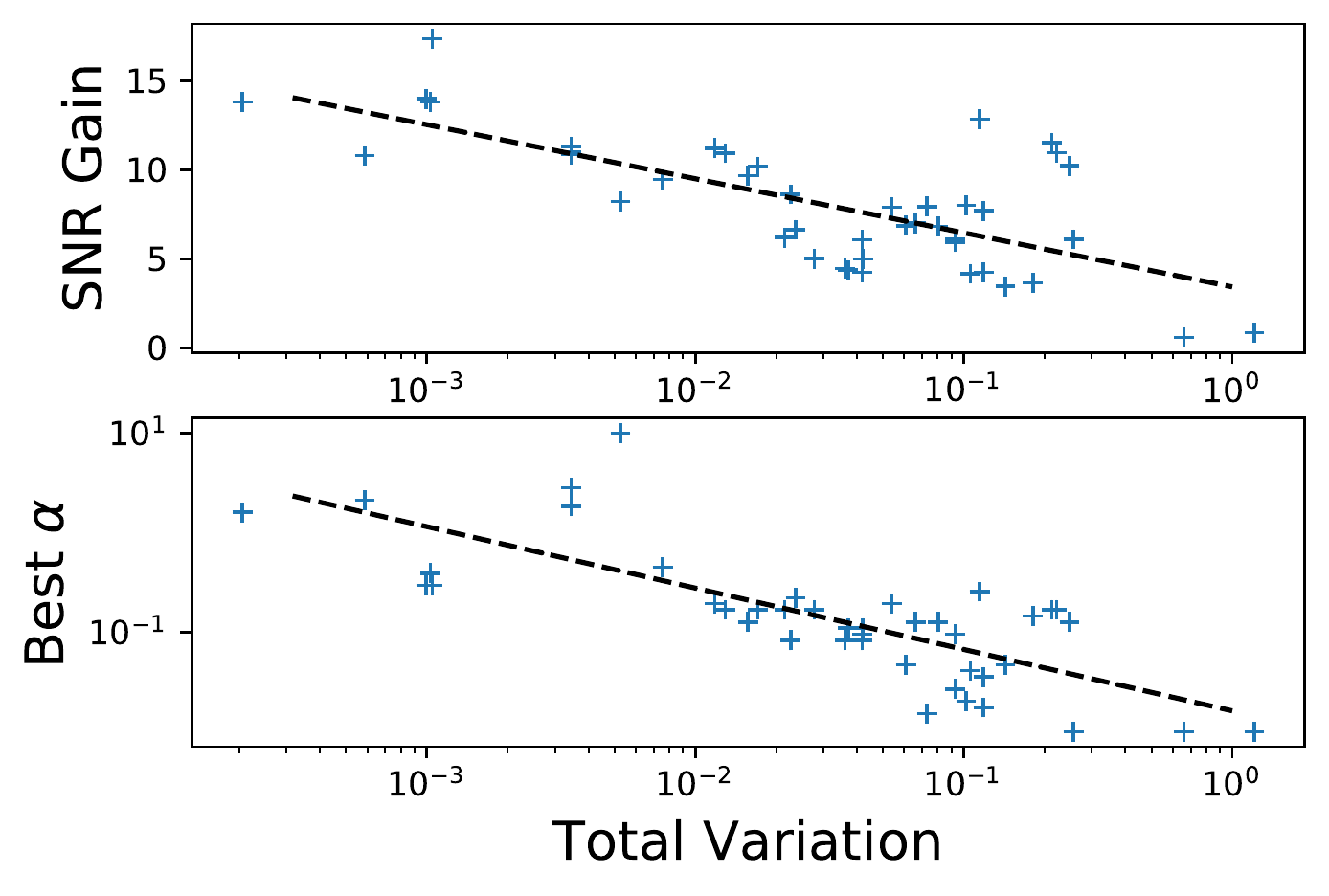}

    %\caption{Efficacy of denoising approaches as a function of noiseless
    %signal low-pass compressibility and \ac{TV}. Note the $x$-axis is used for
    %both quantities.}
    \caption{\textbf{Top}: SNR Gain of denoising vs.\ \ac{TV}, using best
    $\bm{h}_\alpha$ for each network \textbf{Bottom}: Best $\alpha$ for each
    network vs.\ TV. Dashed lines show log-linear and log-log regression fits
    indicating overall inverse correlation.}
    
    \label{fig:TV_vs_SNR}

\end{figure}

Two approaches to the denoising problem are considered here.  The first is to
use the $99.9\%$ low-pass compression threshold for a graph low-pass filter
$\bm{h}_{LP}$, where $h_{LP_k}=1$ for harmonics below the threshold and
$h_{LP_k}=0$ for those above it.  The second denoising approach uses a graph
filter $\bm{h}_\alpha$ determined by \ac{TV} regularization approach
parameterized by $\alpha$, where $\bm{h}_{\alpha_k}=(1+2\alpha\lambda_k)^{-1}$
for the eigenvalues $\lambda_k$ of each network's $\bm{L}^{(|u|)}$
\cite{stankovic2019understanding}.
To assess the efficacy of these filters, for each of the power systems and
default bus voltages $\bm{V}$, we performed a series of Monte Carlo simulations
on these two forms of denoising.  For each default voltage signal $\bm{V}$, we
added a white noise signal $\bm{n}$ so that the expected signal-to-noise ratio
(SNR) was $20$~dB, and then performed the denoising procedure for $\bm{h}_{LP}$
and $\bm{h}_{\alpha}$ for 50 logarithmically spaced $\alpha\in[.01,10]$.  This
procedure was repeated for 25 independent noise perturbations per network, and
the improvement (possibly negative) in output SNR was computed.
Figure~\ref{fig:TV_vs_SNR} (top) shows the average improvement in SNR across
these random samples for the best (over $\alpha$)
$\bm{h}_\alpha$ with the bottom panel showing the best $\alpha$ for that
network.  We see that \ac{TV} correlates with these two quantities quite well
(SNR gain: $r_s=-0.57$, $p=6.7\times10^{-5}$, best $\alpha$: $r_s=-0.75$,
$p=1.0\times10^{-8}$) indicates a strong inverse correlation between the
\ac{TV} of the noiseless signal and improvement in SNR by the denoising
procedure.  Furthermore, we find that the best $\bm{h}_\alpha$ improved the SNR
by about $1.3~dB$ more on average than $\bm{h}_{LP}$, although other compression
ratios may perform better.

\begin{figure}[h]

    \centering

    \includegraphics[width=.9\columnwidth]{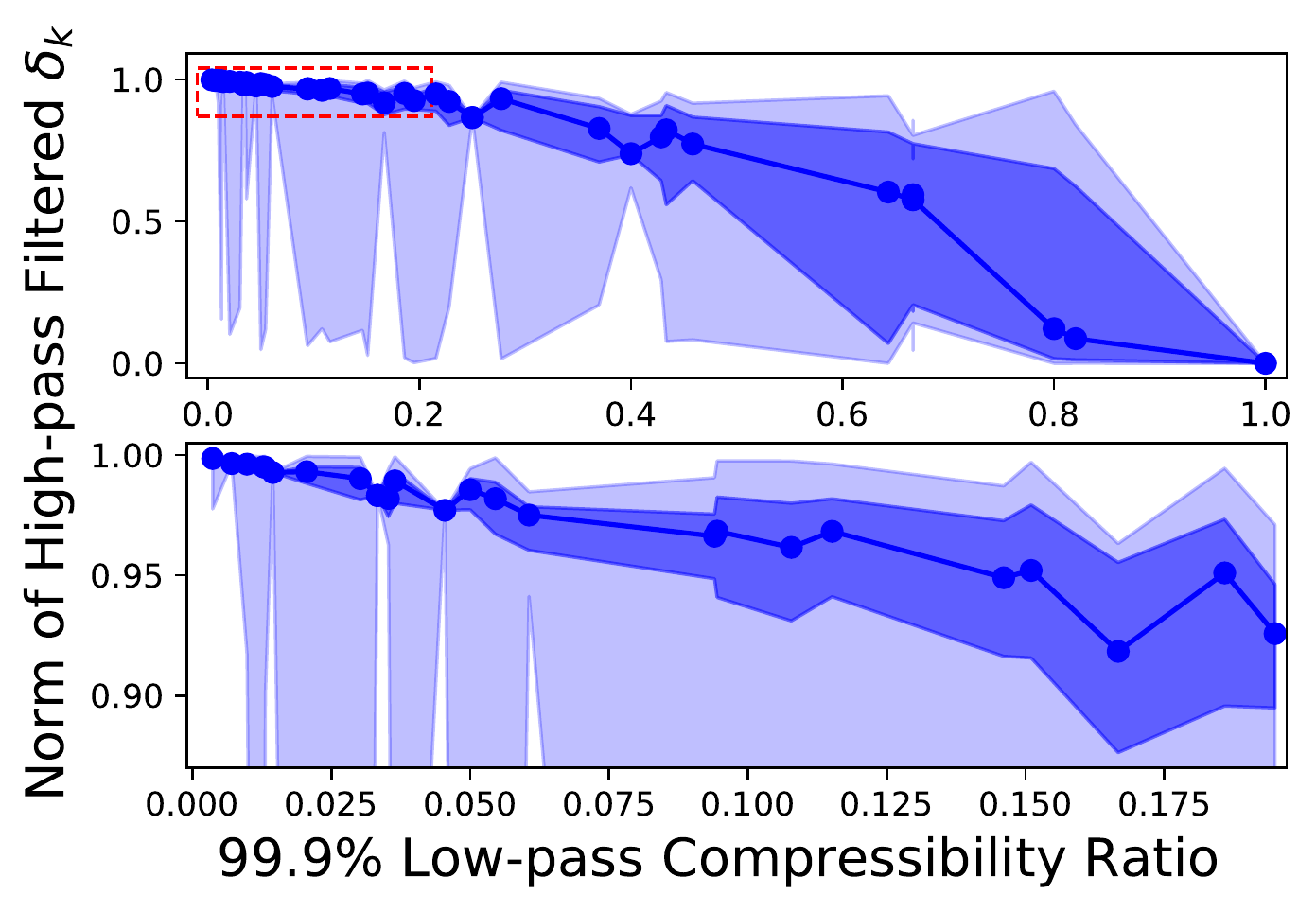}

    \caption{\textbf{Top}: Norm of $\bm{\delta}_k$ filtered using
    $\bm{h}_{HP}$. Line is the median for each network, darker shaded region is
    the inner 50\% quantile, lighter shaded region is the full range.
    \textbf{Bottom}: Zoomed-in view of dashed rectangle from top panel.
    } \label{fig:FDI_energy}

\end{figure}

Next, we consider how the \ac{GSP} model assumptions relate to \ac{FDI}
detection. The approach of \cite{hasnat2020detection} is based on the high-pass
filtering of a signal $\bm{f}$ perturbed by noise $\bm{n}$ and an injected
graph signal $\bm{e}$. Thresholding on the norm of the filtered signal is used
to determine the presence of an \ac{FDI} attack.  To assess how this approach
might work with respect to the systems here, we define a high-pass filter
$\bm{h}_{HP}=1-\bm{h}_{LP}$, using the low-pass filters derived from the 99.9\%
compressibility thresholds above.  We then apply this filter to graph signals
$\bm{\delta}_{k}$ where $\bm{\delta}_k=1$ on $k^{th}$vertex and 0 elsewhere.
This measures how much injected signal will contribute to the detection
threshold (i.e., the closer the filtered norm is to 1, the more detectable an
\ac{FDI} on that vertex will be).  Figure~\ref{fig:FDI_energy} shows the median
norm of the filtered $\bm{\delta}_k$ for each network (sorted by 99.9\%
low-pass compressibility ratio), along with notions of the spread of the
resulting distributions of the norms.  There is a strong correlation between
the median and compressibility ($r_s=-0.99$, $p=1.7\times10^{-34}$), but it is
worth noting that there is substantial variability for many of the networks.
This indicates that certain buses are more susceptible to \ac{FDI} attacks than
others.  

\section{Water Distribution Systems}\label{sec:water}
Unlike power systems, water systems do not appear to have the variety of
\ac{GSP}-motivated analysis in the literature. The exception appears to be the
work in \cite{wei2019optimal,wei2019monitoring} which focuses on the spread of
pollutants 
%analysis of pollutant spread 
in water distribution systems. The authors note that the graph Fourier
structure of the pollutant ``signal'' is not sparse or compressible with
respect to the Laplacian of the unweighted network model.  
%This agrees with \ac{GSP} theory, as we would expect the local spread of a
%contaminant to initially be quite spectrally dense in the graph Fourier domain
%as the contaminant is localized to only a few nodes.  
This lack of signal structure in the graphical Fourier domain led to the
development of so-called ``data-driven'' approaches to designing \ac{GFT}s that
induce the desired compressibility based on a collection of observed data
\cite{wei2019optimal,wei2019monitoring}.  However, these techniques more
closely resemble the PCA-based analysis discussed in Sec.~\ref{sec:sig_model}
than \ac{GSP}, and do account for the structure of the underlying
network.

We conjecture that part of the reason for a lack of \ac{GSP} analysis in this
domain is the absence of canonical relations between signals on the graph
vertices and flows on the graph edges, unlike the power systems case.
There is, however, an analogy between hydraulic and electric circuits that
associates flow-volume through pipes with current through branches, and
pressure differences between junctions with voltage differences between buses.
Unlike Ohm's law between voltage and current, the hydraulic circuit equation
for pressure loss is generally modeled as nonlinear and include a number of
additional physical considerations that make the equations less tractable than
in power systems. Water distribution systems can be modeled as an undirected
graph $\mathcal{G}=(\mathcal{V},\mathcal{E})$, where the vertices are junctions
with head pressure, $H_k$, and the edges are pipes with water flow rate,
$q_{k,\ell}$, representing fluid flow from junction $k$ to $\ell$.  Analogous
to Kirchhoff's current law, the net flow rate of fluids into and out of a
junction can be assumed to be zero barring any exogenous demand $D_k$
(analogous to a current source), so for each junction $k$,
\begin{equation}
    \sum_{(k,\ell)\in \mathcal{E}} q_{k,\ell}=D_k
\end{equation}
which is essentially identical to Kirchhoff's current law.  Hydraulic systems
also satisfy an analogue to Kirchhoff's voltage law, where the net pressure
change over any closed loop is zero (with pumps replacing voltage sources).
However, as noted above, the relation between head loss and water flow is not
generally modeled well by a linear relation like Ohm's law.
A common approximation of the head pressure loss (in $m$) between junctions
(and the one used in simulations here \cite{klise2018overview}) is the
Hazen-Williams headloss formula \cite{rossman2000epanet}:
\begin{equation}\label{eq:HW}
    H_\ell-H_k=\text{sgn}(q_{k,\ell})
    10.667C_{k,\ell}^{-1.852}d_{k,\ell}^{-4.871}L_{k,\ell}|q_{k,\ell}|^{1.852}\,,
\end{equation}
where $C_{k,\ell}$, $d_{k,\ell}$, $L_{k,\ell}$, $q_{k,\ell}$ are the roughness
coefficient (unitless), diameter (in $m$), length (in $m$), and flow rate (in
$m^3/s$) of the pipe connecting junctions $k$ and $\ell$, respectively, and
$\text{sgn}$ is the signum function.
There is also a linear version of the pressure loss equation, called the
Hagen-Poiseuille equation which relates the head loss to the flow rate and pipe
length and diameter via \begin{equation}\label{eq:HP}
    H_\ell-H_k\propto \frac{L_{k,\ell}}{d_{k,\ell}^4}q_{k,\ell}
\end{equation}
where we have ignored a number of constants that will apply to all pipes
identically, and thus act as a global scaling.
% in any \ac{GSP} analysis.

\begin{figure}[h]

    \centering

    \includegraphics[width=.9\columnwidth]{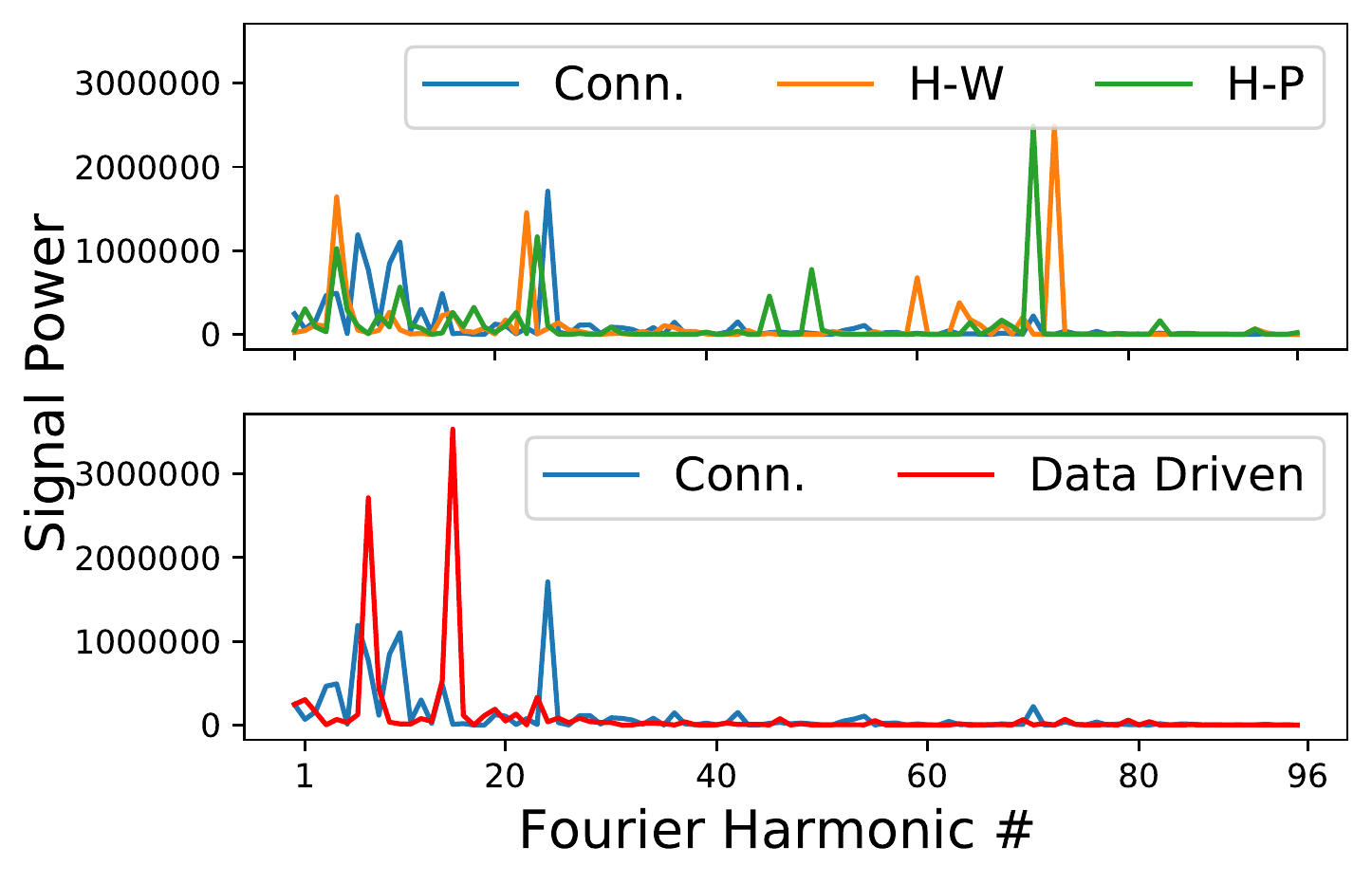}

    \caption{Total signal power of head loss at each junction in the graph
    Fourier domain ($0^{th}$ harmonic omitted for scaling), using different
    weighted network models. \textbf{Top}: \textbf{Conn.}-unweighted network
    using only connectivity, \textbf{H-W}-weights set to the inverse of the
    Hazen-Williams coefficients, \textbf{H-P}-weights set to the inverse of the
    Hagen-Poiseuille coefficients. None of the weighted networks produces
    especially low-pass or sparse graph signals. \textbf{Bottom}:
    \textbf{Conn.}-unweighted network using only connectivity, \textbf{Data
    Driven}-a data-driven approach designed to assign edge weights that produce
    low-pass characteristics.} 
    
    \label{fig:water_combined}

\end{figure}

Using the package \texttt{pyWNTR} \cite{klise2018overview} we simulated the
\texttt{EPANET} \cite{rossman2000epanet} example network 3 (97 junctions, 119
pipes), which simulated a time series of hydraulic simulations based on a
simulated demand model (673 graph signals over 168 hours).
Figure~\ref{fig:water_combined} (top) shows the \ac{GFT} power spectrum of the
simulated signals for Laplacians computed using the unweighted connectivity as
well as weights from \eqref{eq:HW} and \eqref{eq:HP}.  We see that none of
these approaches produce signals that are especially compressible or smooth on
average (outside of the dominant $0^{th}$ harmonic).  An open question is
the existence of a principled, data-driven approach that also accounts for the
underlying connectivity of the network.  To this end, we can use random search
to find weights that improve the overall compressibility of the set of signals,
see Figure~\ref{fig:water_combined} (bottom), but we leave a more principled
approach to future research.  This indicates that it may be possible to merge
purely data-driven approaches with \ac{GSP}-motivated approaches to create
network models driven by the underlying physics to create \ac{GFT}s where the
graph signals strongly meet the model assumptions, producing better results for
the \ac{GSP} technique.

%\begin{figure}[h]
%
%    \centering
%
%    \includegraphics[width=.9\columnwidth]{figs/WNTR_GFT_compare}
%
%    \caption{Total signal power of head loss at each junction in the graph
%    Fourier domain ($0^{th}$ harmonic omitted for scaling), using different
%    weighted network models: \textbf{Conn.}-unweighted network using only
%    connectivity, \textbf{H-W}-weights set to the inverse of the Hazen-Williams
%    coefficients, \textbf{H-P}-weights set to the inverse of the
%    Hagen-Poiseuille coefficients. None of the weighted networks produces
%    especially low-pass graph signals.}
%    \label{fig:WNTR_GFT_compare}
%
%\end{figure}
%
%
%\begin{figure}[h]
%
%    \centering
%
%    \includegraphics[width=.9\columnwidth]{figs/WNTR_GFT_compressed}
%
%    \caption{Comparison of signal power in the graph Fourier domian for
%    transforms based on the unweighted graph (\textbf{Conn.}) and a data-driven
%    approach designed to assign edge weights that produce low-pass
%    characteristics in the graph signals defined by head loss (\textbf{Data
%    Driven}).}
%    \label{fig:WNTR_GFT_compressed}
%
%\end{figure}

%\section{Other Infrastructure Applications}
%%
%Traffic monitoring \cite{chen2016monitoring}, pollutant sensor networks for
%smart cities \cite{jablonski2017graph}

\section{Conclusion}
In conclusion, we have demonstrated the importance of understanding \ac{GSP}
model assumptions in the context of infrastructure resilience applications.
We emphasize that this work should not be interpreted as a definitive approach
on the chosen techniques, rather it should be interpreted more pedagogically,
as a case study analysis in a particular choice of \ac{GFT} and signal model,
similar to any analysis that must be performed in any practical, real-world
application of these techniques. 
We found that system size and distributed generation capability were reasonably
strong correlates to relevant \ac{GSP} metrics and ultimately performance of
considered \ac{GSP} techniques.  Thus, we note that the variability introduced
by distributed generation (e.g., renewables) in relatively small networks may
limit \ac{GSP} techniques in micro-grid applications, and in any event should
be analyzed carefully to make sure the graph signals meet the required
assumptions.  

Despite the observed correlations, there does appear to be considerable
variation in both the metrics and performance for infrastructure systems of
similar size and scope, pointing to a need for further analysis.
Other than system size and distributed generation capability, we did not
identify any particular characteristics of the networks that generated such
variation in the metrics and results, even among networks of similar size.
There are many operational and graph-centric metrics that can be explored
\cite{holmgren2006using,pagani2013power, pagani2020quantifying,pagano2019water}
to seek further insight, which we leave as future work.  Similarly, given the
usage of graph-theoretic metrics in resilience analysis of infrastructure, we
conjecture that \ac{GSP}-motivated metrics such as \ac{TV} can provide relevant
metrics that capture the state of the system over time, in a way that static
network metrics cannot.

Beyond the AC power-flow derived graph model and $\bm{L}^{(|u|)}$-based
\ac{GFT} used here, additional combinations of graph models (such as those
discussed in Sec.~\ref{sec:power}) and \ac{GFT}s may be more effective for
specific combinations of system and application.  The results of
Sec.~\ref{sec:water} indicate that data-driven approaches may be able to
leverage graphical structure to produce \ac{GFT}s that induce desirable signal
structure, and this generates additional options.  Understanding how the choice
of graph model and \ac{GFT} impacts performance of a given \ac{GSP} technique
on a given system is an open question that should be studied to motivate new
techniques as well as reduce future implementation and development efforts in
real-world systems.

\bibliographystyle{IEEEtran}
\bibliography{references}

\end{document}